\documentstyle[referee]{laa}

\begin{document}
\thesaurus{07  
	   (07.09.1; 
	    07.13.1;  
	   )}

\title{Radiation and Dynamics of Dust Particle}
\author{J.~Kla\v{c}ka}
\institute{Astronomical Institute,
   Faculty of Mathematics, Physics, and Informatics \\ Comenius University,
   Mlynsk\'{a} dolina, 842~48 Bratislava, Slovak Republic \\
   E-mail: klacka@fmph.uniba.sk}
\date{}
\maketitle

\begin{abstract}
Relativistically covariant form of equation of motion for arbitrarily shaped
dust particle (neutral in charge) under the action of electromagnetic
radiation is derived -- emission, scattering and absorption of radiation
is considered.
The result is presented in the form of optical quantities used in optics
of dust particles. The obtained equation of motion represents a generalization
of the Poynting-Robertson (P-R) effect, which is standardly used in orbital
evolution of dust particles in astrophysics.

Simultaneous action of electromagnetic radiation and
gravitational fields of the central body -- star -- on the motion
of the particle is discussed.


\end{abstract}

\section{Introduction}
The first case closely connected with relativistic equation of motion for a
particle under the action of electromagnetic radiation was presented by
Einstein (1905), who calculated the change of energy and light pressure for
plane mirror. Later on, Robertson (1937) has presented equation of motion for
perfectly absorbing spherical particle -- this case has been applied to
astronomical situations for several decades as the well-known
Poynting-Robertson (P-R) effect.

It is inevitable to have in disposal general equation of motion.
Equation of motion for all types of particles, not only for the special
cases treated by the above cited authors. Real particles scatter radiation
in a more complicated manner and particles of various optical properties occur.
Moreover, thermal emission has to be included, in order to be the effect of
radiation complete. Thus, this paper generalizes the results presented in
Kla\v{c}ka (2000a, 2000b, 2001, 2002a) and Kla\v{c}ka and Kocifaj (2001), in
which also results of detailed numerical calculations for (sub)micron sized
dust particles in the Solar System may be found -- motion of real particles
may significantly differ from that corresponding to the P-R effect.

\section{Proper reference frame of the particle -- stationary particle}
The term ``stationary particle'' will denote particle which does
not move in a given inertial frame of reference.
Primed quantities will denote quantities measured in the
proper reference frame of the particle.

The flux density of photons scattered into an element of solid angle
$d \Omega ' = \sin \theta ' ~ d \theta ' ~ d \phi '$
is proportional to  $p' ( \theta ', \phi ') ~ d \Omega '$, where
$p' ( \theta ', \phi ')$ is ``phase function''.
Phase function depends
on orientation of the particle with respect to the direction of the
incident radiation and on the particle characteristics;
angles $\theta '$, $\phi '$ correspond to the direction (and orientation)
of travel of the scattered radiation, $\theta '$ is polar angle
and it equals zero for the case of the travel of the ray in the orientation
identical with the unit vector $\vec{e}_{1} '$ of the incident radiation.
The phase function fulfills the condition
\begin{equation}\label{1}
\int_{4 \pi} p' ( \theta ', \phi ')~ d \Omega ' = 1 ~.
\end{equation}

The momentum of the incident beam of photons which is lost in the process
of its interaction with the particle is proportional to the cross section
$C'_{ext}$ (extinction). The part proportional to $C'_{abs}$ (absorption)
is completely lost and the part proportional to
$C'_{ext} ~-~ C'_{abs} = C'_{sca}$ (scattering) is again reemitted.
The differential scattering cross section $d C'_{sca} / d \Omega '$
$\equiv$ $C'_{sca} ~ p' ( \theta ', \phi ' )$ depends on the
polarization state of the incident light as well as on the incidence
and scattering directions (e. g., Mishchenko et al. 2002).

The momentum (per unit time) of the scattered photons into an element
of solid angle $d \Omega '$ is
\begin{equation}\label{2}
d \vec{p'}_{sca} = \frac{1}{c} ~ S' ~ C'_{sca} ~
		   p' ( \theta ', \phi ')~ \vec{K'} ~ d \Omega ' ~,
\end{equation}
where unit vector in the direction of scattering is
\begin{equation}\label{3}
\vec{K'} = \cos \theta '~\vec{e}_{1} ' ~+~
		 \sin \theta ' ~ \cos \phi ' ~ \vec{e}_{2} ' ~+~
		 \sin \theta ' ~ \sin \phi ' ~ \vec{e}_{3} ' ~.
\end{equation}
$S'$ is the flux density of radiation energy. The system of unit vectors
used on the right-hand side of the last equation forms an orthogonal basis.
The total momentum (per unit time) of the scattered photons is
\begin{equation}\label{4}
\vec{p'}_{sca} = \frac{1}{c} ~ S' ~ C'_{sca} ~ \int_{4 \pi} ~
		 p' ( \theta ', \phi ')~ \vec{K'} ~ d \Omega ' ~.
\end{equation}

The momentum (per unit time) obtained by the particle due
to the interaction with radiation is
\begin{equation}\label{5}
\frac{d~ \vec{p'}}{d~ t'} = \frac{1}{c} ~ S' ~ \left \{
			      C'_{ext} ~\vec{e}_{1} '
			      ~-~ C'_{sca} ~ \int_{4 \pi} ~
			      p' ( \theta ', \phi ')~ \vec{K'} ~
			      d \Omega ' \right \} ~.
\end{equation}

As for the energy, we suppose that it is conserved: the energy (per unit time)
of the incoming radiation, equals to the energy (per unit time)
of the outgoing radiation (after interaction with the particle).
We will use the fact that $t' = \tau$, where $\tau$ is proper time.

For the sake of brevity, we will use dimensionless efficiency factors $Q'_{x}$
instead of cross sections $C'_{x}$:
$C'_{x} = Q'_{x} ~ A'$, where $A'$ is geometrical
cross section of a sphere of volume equal to the volume of the
particle. Equation (5) can be rewritten to the form
\begin{eqnarray}\label{6}
\frac{d ~\vec{p'}}{d~ \tau} &=& \frac{1}{c} ~ S'~A'~ ~ \left \{ \left [
	     Q'_{ext} ~-~ < \cos \theta'> ~ Q'_{sca} \right ] ~
	     \vec{e}_{1} ' ~+~ \right .
\nonumber \\
& &  \left . \left [ ~-~ < \sin \theta' ~ \cos \phi ' > ~ Q'_{sca}
	     \right ] ~ \vec{e}_{2} ' ~+~
	     \left [ ~-~ < \sin \theta' ~ \sin \phi ' > ~ Q'_{sca}
	     \right ] ~ \vec{e}_{3} ' \right \} ~.
\end{eqnarray}

Summarizing important equations, we can write them in a short form
\begin{equation}\label{7}
\frac{d~ \vec{p'}}{d~ \tau} = \frac{1}{c} ~ S'~A'~ ~ \left \{
	     Q_{1} ' ~ \vec{e}_{1} ' ~+~ Q_{2} '~ \vec{e}_{2} '
	     ~+~ Q_{3} '~ \vec{e}_{3} ' \right \} ~; ~~
\frac{d ~E'}{d~ \tau} = 0 ~.
\end{equation}

\section{Stationary frame of reference}
By the term ``stationary frame of reference'' we mean a frame of reference
in which particle moves with a velocity vector $\vec{v} = \vec{v} (t)$.
The physical quantities measured in the stationary frame of reference
will be denoted by unprimed symbols.

Our aim is to derive equation of motion for the particle in the
stationary frame of reference.
We will use the fact that we know
this equation in the proper frame of reference -- see Eqs. (7).
We have to use Lorentz transformation for the purpose of making
transformation from proper frame of reference to stationary frame
of reference.

If we have a four-vector $A^{\mu} = ( A^{0}, \vec{A} )$, where
$A^{0}$ is its time component and $\vec{A}$ is its spatial component,
generalized special Lorentz transformation yields
\begin{eqnarray}\label{8}
A^{0 '}  &=& \gamma ~ ( A^{0} ~-~ \vec{v} \cdot \vec{A} / c ) ~,
\nonumber \\
\vec{A} ' &=& \vec{A} ~+~ [ ( \gamma ~-~ 1 ) ~ \vec{v} \cdot \vec{A}  /
	      \vec{v} ^{2} ~-~ \gamma ~ A^{0} / c ] ~ \vec{v}  ~,
\end{eqnarray}
where $\gamma = 1 / \sqrt{1~-~\vec{v} ^{2} / c ^{2} }$.
The inverse generalized special Lorentz transformation is
\begin{eqnarray}\label{9}
A^{0}  &=& \gamma ~ ( A^{0 '} ~+~ \vec{v} \cdot \vec{A} ' / c ) ~,
\nonumber \\
\vec{A}  &=& \vec{A} ' ~+~ [ ( \gamma ~-~ 1 ) ~ \vec{v} \cdot \vec{A} ' /
	      \vec{v} ^{2} ~+~ \gamma ~ A^{0 '} / c ] ~ \vec{v}  ~.
\end{eqnarray}

Well-known four-vector is four-momentum:
$p^{\mu} = ( p^{0}, \vec{p} ) \equiv ( E / c, \vec{p} )$.

\subsection{Transformation of equation of motion}
Eqs. (9) yield for our case represented by the second of Eqs. (7):
\begin{eqnarray}\label{10}
\frac{dE}{d \tau} &=& \gamma ~ \vec{v} \cdot \frac{d \vec{p}'}{d \tau} ~,
\nonumber \\
\frac{d \vec{p}}{d \tau} &=& \frac{d \vec{p}'}{d \tau} ~+~
		   \left ( \gamma ~-~ 1 \right ) ~
		   \left ( \vec{v} \cdot \frac{d \vec{p}'}{d \tau} \right )~
		   \frac{\vec{v}}{\vec{v}^{2}} ~.
\end{eqnarray}

\subsection{Flux density of radiation energy}
Let us consider that a beam of incident photons is characterized by
flux density of radiation energy $S' = n' h \nu ' c$, or
$S = n h \nu  c$, where $n'$ and $n$ are concentrations of photons
in the corresponding reference frames. The incident photon is characterized
by four-momentum $p^{\mu}$ $=$ $(h \nu$, $h \nu \vec{e}_{1})$. Application
of Eqs. (8) to the four-momentum of the photon yields:
\begin{eqnarray}\label{11}
\nu ' &=& \nu  ~w_{1} ~,
\nonumber \\
\vec{e}_{1} ' &=& \frac{1}{w_{1}}  ~ \left \{ \vec{e}_{1} ~+~
		 \left [ \left ( \gamma ~-~ 1 \right ) ~
		 \vec{v} \cdot \vec{e}_{1}  /
		 \vec{v} ^{2} ~-~ \gamma / c \right ] ~ \vec{v} \right \} ~,
\end{eqnarray}
where abbreviation
$w_{1}$ $\equiv$ $\gamma ~ ( 1 ~-~ \vec{v} \cdot \vec{e}_{1} / c )$
is used. Continuity equation $\partial _{\mu} j^{\mu} =$ 0, where
$j^{\mu} =$ $(c n$, $c n \vec{e}_{1})$ is four-vector of current density,
yields $n' = n w_{1}$. Inserting this result together with the first of Eqs. (11)
into the flux density of radiation energy $S'$, we obtain
\begin{equation}\label{12}
S' = w_{1}^{2} ~S ~.
\end{equation}

\subsection{Transformation of unit vectors}
We have already found transformation of the vector $\vec{e}_{1}$. We have
to found also transformations of vectors $\vec{e}_{2}$ and $\vec{e}_{3}$.
Physics of the corresponding transformation is that the vectors
$\vec{e}_{2}$ and $\vec{e}_{3}$ describe directions of propagation of
light after interaction with the particle (see Kla\v{c}ka 2000a for more
details). Thus, aberration of light exists for each of these unit vectors.
Considerations analogous to those for vector $\vec{e}_{1}$ immediately yield
\begin{eqnarray}\label{13}
\vec{e}_{j} ' &=& \frac{1}{w_{j}}  ~ \left \{ \vec{e}_{j} ~+~
		  \left [ \left ( \gamma ~-~ 1 \right ) ~
		  \vec{v} \cdot \vec{e}_{j}  /
		  \vec{v} ^{2} ~-~ \gamma / c \right ] ~ \vec{v} \right \} ~,
		  ~~ j = 1, 2, 3 ~,
\nonumber \\
w_{j} &\equiv& \gamma ~ ( 1 ~-~ \vec{v} \cdot \vec{e}_{j} / c ) ~.
\end{eqnarray}

\subsection{Equation of motion}
Inserting Eqs. (7), (12) and (13) into Eqs. (10),
one easily obtains
\begin{eqnarray}\label{14}
\frac{d ~E}{d~ \tau} &=&  \sum_{j=1}^{3} ~Q_{j} ' ~ w_{1}^{2} ~S ~A' ~
	  \left ( 1 / w_{j}  ~-~ \gamma \right ) ~,
\nonumber \\
\frac{d ~\vec{p}}{d~ \tau} &=& \sum_{j=1}^{3} ~ Q_{j} ' ~
			       \frac{w_{1}^{2} ~S ~A'}{c^{2}} ~
			       \left ( c ~\vec{e}_{j} / w_{j} ~-~
			       \gamma \vec{v}  \right )  ~.
\end{eqnarray}

Eqs. (14) may be rewritten in terms of four-vectors:
\begin{equation}\label{15}
\frac{d ~p^{\mu}}{d~ \tau} = \frac{w_{1}^{2} ~S ~A'}{c^{2}} ~
			       \sum_{j=1}^{3} ~Q_{j} ' ~
			       \left ( c ~ b_{j}^{\mu} ~-~ u^{\mu}  \right ) ~,
\end{equation}
where $p^{\mu}$ is four-vector of the particle of mass $m$,
$p^{\mu}$ $=$ $m~ u^{\mu}$,
four-vector of the world-velocity of the particle is
$u^{\mu}$ $=$ $( \gamma ~c, \gamma ~ \vec{v} )$.
We have also other four-vectors
\begin{equation}\label{16}
b_{j}^{\mu} = ( 1 / w_{j} , \vec{e}_{j} / w_{j} ) ~, ~~ j= 1, 2, 3 ~.
\end{equation}

It can be easily verified that Eq. (15) yields $d ~m / d~ \tau =$ 0.

As for practical calculations we introduce:
\begin{eqnarray}\label{17}
b_{j}^{0} &\equiv& 1 / w_{j} = \gamma ~( 1 ~+~ \vec{v} \cdot \vec{e}'_{j} ~/~c ) ~, ~~ j= 1, 2, 3,
\nonumber \\
\vec{b_{j}} &\equiv& \vec{e_{j}} ~/~ w_{j} = \vec{e}'_{j} ~+~ [ ( \gamma ~-~ 1 )~
		   \vec{v} \cdot \vec{e}'_{j} ~/~ \vec{v}^{2} ~+~
		   \gamma ~/~c ] ~ \vec{v} ~.
\end{eqnarray}

Within the accuracy to the first order in $\vec{v} / c$, Eq. (15) yields
\begin{equation}\label{18}
\frac{d~ \vec{v}}{d ~t} =  \frac{S ~A'}{m~c} ~ \sum_{j=1}^{3} ~Q_{j} '
			  ~\left [  \left ( 1~-~ 2~
	      \vec{v} \cdot \vec{e}_{1} / c ~+~
	      \vec{v} \cdot \vec{e}_{j} / c \right ) ~ \vec{e}_{j}
	      ~-~ \vec{v} / c \right ]	~.
\end{equation}

Eq. (15) yields as special cases the situations discussed in Einstein (1905)
and Robertson (1937) -- Robertson's case is obtained simply putting
$Q'_{1} =$ 1, $Q'_{2} = Q'_{3} =$ 0, Einstein's results require a little more
calculations.

\subsection{Heuristic derivation}
Since we completely understand physics of Eq. (15), we are able to present
short simple derivation of Eq. (18), now.
We have: \\
i) $d \vec{v'} / dt = [ S' A' / ( m c ) ] ~\sum_{i=1}^{3}
	     Q_{i} ' ~ \vec{e}_{i} '$ (see Eq. (7)); \\
ii) $S' = S ( 1 - 2 \vec{v} \cdot \vec{e}_{1} / c )$ (see Eq. (12)),
due to the change of concentration of photons
$n' = n ( 1 - \vec{v} \cdot \vec{e}_{1} / c )$ and Doppler effect
$\nu' = \nu ( 1 - \vec{v} \cdot \vec{e}_{1} / c )$; \\
iii) $\vec{e}_{j} ' = ( 1 + \vec{v} \cdot \vec{e}_{j} / c ) \vec{e}_{j} -
\vec{v} / c$, $j =$ 1, 2, 3 (aberration of light -- see Eq. (13)). \\
Taking into account these physical phenomena, we finally obtain Eq. (18).

\section{Thermal emission}
If the particle's absolute temperature is above zero, it can emit as well
as scatter and absorb electromagnetic radiation.

\subsection{Proper reference frame of the particle -- stationary particle}
This subsection is based on the presentation by Mishchenko et. al. (2002).

The emission component of the radiation force acting on the particle
of absolute temperature $T'$ is
\begin{equation}\label{19}
\vec{F}'_{e} ( T' ) = -~ \frac{1}{c} ~ \int_{0}^{\infty} ~ d \omega' ~
		      \int_{4 \pi} ~ \hat{\vec{r}}' ~
		      K'_{e} \left ( \hat{\vec{r}}', T', \omega ' \right ) ~
		      d \hat{\vec{r}}' ~,
\end{equation}
where the unit vector $\hat{\vec{r}}' = \vec{r}' / r'$ is given by position vector
$\vec{r}'$ of the observation point with origin inside the particle
(the emitted radiation in the far-field zone of the particle propagates
in the radial direction, i. e., along the unit vector $\hat{\vec{r}}'$),
$\omega '$ is (angular) frequency of radiation,
\begin{equation}\label{20}
K'_{e} \left ( \hat{\vec{r}}', T', \omega ' \right ) =
       I'_{b} \left ( T', \omega ' \right ) ~ \left \{
       K'_{11} \left ( \hat{\vec{r}}', \omega ' \right ) ~-~
       \int_{4 \pi} ~
       Z'_{11} \left ( \hat{\vec{r}}', \hat{\vec{r}}'', \omega ' \right ) ~
		      d \hat{\vec{r}}'' \right \}~,
\end{equation}
and the Planck blackbody energy distribution is given by the well-known relation
\begin{equation}\label{21}
I'_{b} \left ( T', \omega ' \right ) =
       \frac{\hbar ~\omega'^{3}}{4~ \pi ^{3} ~c^{2}}
       \left \{ \exp \left ( \frac{\hbar ~\omega'}{k~T'} \right )
       ~-~ 1 \right \}^{-1} ~.
\end{equation}
The functions $K'_{11}$ and $Z'_{11}$
\begin{eqnarray}\label{22}
K'_{11} &=& \frac{2 ~\pi}{k'_{1}} ~ Im \left ( S'_{11} ~+~ S'_{22} \right ) ~,
\nonumber   \\
Z'_{11} &=& \left ( | S'_{11} | ^{2} ~+~ | S'_{12} | ^{2} ~+~ | S'_{21} | ^{2}
	    ~+~ | S'_{22} | ^{2}  \right ) ~/~ 2 ~,
\end{eqnarray}
are given by elements of the
amplitude matrix $S' ( \hat{\vec{n}}'^{sca}, \hat{\vec{n}}'^{inc} )$:
\begin{equation}\label{23}
\vec{E}'^{sca} ( r' ~ \hat{\vec{n}}'^{sca} ) = \frac{e^{i~k'_{1} ~r'}}{r'} ~
	       S' ( \hat{\vec{n}}'^{sca}, \hat{\vec{n}}'^{inc} ) ~
\vec{E}'^{inc}_{0}~~,~~~ r' \rightarrow \infty ~.
\end{equation}
We assume that the incident field is a plane electromagnetic wave given by
\begin{equation}\label{24}
\vec{E}'^{inc} ( \vec{r}' ) = \vec{E}'^{inc}_{0}~
\exp \left ( i~k'_{1} \hat{\vec{n}}'^{inc} \cdot \vec{r}' \right ) ~~, ~~~
\omega'^{2} = c^{2} ~ | \vec{k}'_{1} |^{2} ~,
\end{equation}
since we consider vacuum as a homogeneous nonabsorbing medium. We use
a couplet ( $\vartheta'$, $\varphi'$ ), where $\vartheta'$ $\in$ $\langle$
$0, \pi$ $\rangle$ is the polar (zenith) angle measured from the positive
$z'-$axis and $\varphi'$ $\in$ $\langle$ $0, 2~ \pi$ $\rangle$ is the
azimuth angle measured from the positive $x'-$axis in the clockwise direction
when looking in the direction of the positive $z'-$axis. Since the medium is
assumed to be nonabsorbing,
$\vec{E}'$ $=$ $E'_{\vartheta '}$ $\hat{\vec{\vartheta}}'$ $+$
		       $E'_{\varphi '}$ $\hat{\vec{\varphi}}'$,
where $E'_{\vartheta '}$ $\hat{\vec{\vartheta}}'$ lies in the meridional plane
(i. e., the plane through $\hat{\vec{n}}'$ and the $z'-$axis), whereas the
component $E'_{\varphi '}$ $\hat{\vec{\varphi}}'$ is perpendicular to this plane;
$\hat{\vec{\vartheta}}'$ and $\hat{\vec{\varphi}}'$ are the corresponding unit
vectors such that $\hat{\vec{n}}'$ $=$ $\hat{\vec{\vartheta}}'$ $\times$
$\hat{\vec{\varphi}}'$:
\begin{eqnarray}\label{25}
\vec{E}'^{inc}_{0} &=& E'^{inc}_{0 \vartheta '} ~ \hat{\vec{\vartheta}}'^{inc} ~+~
		       E'^{inc}_{0 \varphi '} ~ \hat{\vec{\varphi}}'^{inc} ~,
\nonumber   \\
\vec{E}'^{sca} &=& E'^{sca}_{\vartheta '} ~ \hat{\vec{\vartheta}}'^{sca} ~+~
		       E'^{sca}_{\varphi '} ~ \hat{\vec{\varphi}}'^{sca} ~~, ~~~
	  ( \hat{\vec{r}'} \cdot \vec{E}'^{sca} ( \vec{r} ' ) = 0 ~~,~~~
	  r' \rightarrow \infty ) ~.
\end{eqnarray}

\subsection{Stationary frame of reference}
Using orthonormal vectors $\vec{e}'_{1}$, $\vec{e}'_{2}$ and $\vec{e}'_{3}$
defined by Eq. (3), we can write Eq. (19) in the form
\begin{equation}\label{26}
\vec{F}'_{e} ( T' ) = \sum_{i=1}^{3} ~ \left [ \vec{F}'_{e} ( T' ) \cdot
		      \vec{e}'_{i} \right ] ~ \vec{e}'_{i} ~.
\end{equation}
Using four-vectors defined by Eq. (16), we can immediately write
\begin{equation}\label{27}
\left ( \frac{d ~p^{\mu}}{d~ \tau} \right ) _{e} =
			 \sum_{j=1}^{3} ~Q_{ej} ' ~ b_{j}^{\mu} ~,
\end{equation}
where
\begin{equation}\label{28}
Q_{ej} ' =  -~ \vec{e}'_{j} \cdot \frac{1}{c} ~ \int_{0}^{\infty} ~ d \omega' ~
	       \int_{4 \pi} ~ \hat{\vec{r}}' ~
	       K'_{e} \left ( \hat{\vec{r}}', T', \omega ' \right ) ~
		      d \hat{\vec{r}}' ~~, ~~~ j = 1, 2, 3 ~.
\end{equation}
As a consequence, Eqs. (16) and(27) yield
\begin{equation}\label{29}
\left ( \frac{d ~m}{d~ \tau} \right ) _{e} =
			 \sum_{j=1}^{3} ~Q_{ej} ' ~/~c ~.
\end{equation}
Mass of the particle decreases due to the thermal emission, alone.

\section{Electromagnetic interaction -- total}

\subsection{Proper reference frame of the particle -- stationary particle}
We can write, on the basis of Eqs. (7), (26) and (28):
\begin{equation}\label{30}
\frac{d~ \vec{p'}}{d~ \tau} = \sum_{j=1}^{3} \left ( \frac{S'~A'}{c} ~
	     Q_{j} ' ~+~ Q_{ej} ' \right ) ~\vec{e}_{j} ' ~;~~
\frac{d ~E'}{d~ \tau} = 0 ~.
\end{equation}

\subsection{Stationary frame of reference}
On the basis of Eqs. (12), (16) and (30) we obtain
\begin{equation}\label{31}
\frac{d ~p^{\mu}}{d~ \tau} = \sum_{j=1}^{3} \left (
		  \frac{w_{1}^{2} ~S ~A'}{c} ~Q_{j} ' ~+~ Q_{ej} '  \right ) ~
			       \left ( b_{j}^{\mu} ~-~ u^{\mu} ~/~c  \right ) ~.
\end{equation}
As a consequence, $d~m~/~d ~\tau =$ 0.

\section{Gravitational and electromagnetic radiation fields}
Let us consider orbital evolution of real dust particle in the Solar System,
under action of gravitational and electromagnetic radiation fields of the Sun.
We can write (also formulation presented in Kla\v{c}ka 2002b is used)
\begin{eqnarray}\label{32}
\frac{d~ \vec{v}}{d ~t} &=& -~ \frac{G~M_{\odot}}{r^{2}} ~ \vec{e}_{1} ~+~
	      \frac{G~M_{\odot}}{r^{2}} ~
	      \sum_{j=1}^{3} ~\beta_{j} ~\left [  \left ( 1~-~ 2~
	      \frac{\vec{v} \cdot \vec{e}_{1}}{c} ~+~
	      \frac{\vec{v} \cdot \vec{e}_{j}}{c} \right ) ~ \vec{e}_{j}
	      ~-~ \frac{\vec{v}}{c} \right ]  ~+~
\nonumber   \\
& &	      \sum_{j=1}^{3} ~Q'_{ej} ~\left [  \left ( 1~+~
	      \frac{\vec{v} \cdot \vec{e}_{j}}{c} \right ) ~ \vec{e}_{j}
	      ~-~ \frac{\vec{v}}{c} \right ]  ~,
\end{eqnarray}
where
\begin{eqnarray}\label{33}
\vec{e}_{j} &=& ( 1 ~-~ \vec{v} \cdot \vec{e'}_{j} / c ) ~ \vec{e'}_{j}  ~+~
	      \vec{v} / c ~, ~~j = 1, 2, 3 ~,
\nonumber   \\
\beta_{1} &=& \frac{\pi ~R_{\odot}^{2}}{G~M_{\odot}~m~c}
	      \int_{0}^{\infty} ~B_{\odot} ( \lambda ) \left \{
	      C'_{ext} ( \lambda / w ) ~-~ C'_{sca} ( \lambda / w ) ~
	      g'_{1} ( \lambda / w ) \right \} ~ d \lambda ~,
\nonumber   \\
\beta_{2} &=& \frac{\pi ~R_{\odot}^{2}}{G~M_{\odot}~m~c}
	      \int_{0}^{\infty} ~B_{\odot} ( \lambda ) \left \{
	      ~-~ C'_{sca} ( \lambda / w ) ~
	      g'_{2} ( \lambda / w ) \right \} ~ d \lambda ~,
\nonumber   \\
\beta_{3} &=& \frac{\pi ~R_{\odot}^{2}}{G~M_{\odot}~m~c}
	      \int_{0}^{\infty} ~B_{\odot} ( \lambda ) \left \{
	      ~-~ C'_{sca} ( \lambda / w ) ~
	      g'_{3} ( \lambda / w ) \right \} ~ d \lambda ~,
\nonumber   \\
w &=& 1~-~\vec{v} \cdot \vec{e}_{1} ~/~ c ~,
\end{eqnarray}
$R_{\odot}$ denotes the radius of the Sun and $B_{\odot} ( \lambda )$ is
the solar radiance at a wavelength of $\lambda$; $G$, $M_{\odot}$, and
$r$ are the gravitational constant, the mass of the Sun, and the distance
of the particle from the center of the Sun, respectively.
The asymmetry parameter vector $\vec{g}'$ is defined by $\vec{g}'$ $=$
$( 1 / C'_{sca} ) \int \vec{n}' ( d C'_{sca} / d \Omega ') d \Omega '$, where
$\vec{n}'$ is a unit vector in the direction of scattering;
$\vec{g}'$ $=$ $g'_{1} ~ \vec{e}'_{1}$ $+$ $g'_{2} ~ \vec{e}'_{2}$ $+$
$g'_{3} ~ \vec{e}'_{3}$, $\vec{e}'_{1}$ $=$
$( 1 ~+~ \vec{v} \cdot \vec{e}_{1} / c ) ~ \vec{e}_{1}$  $-$
$\vec{v} / c$, $\vec{e}'_{i} \cdot \vec{e}'_{j} = \delta _{ij}$.

\section{Conclusion}
We have derived relativistically covariant equation of motion
for dust particle under the action of electromagnetic radiation -- see
Eq. (31). As a generalization requires, Eq. (31) may be reduced to
special cases treated by Einstein (1905), Robertson (1937) and Kla\v{c}ka
(2000a). Application of Eq. (31) to Solar System is represented by Eqs. (32)
and (33). Some other accelerations can be added to the right-hand side of 
Eq. (32) -- e. g., gravitational perturbations of planets, solar wind effect
$[(d \vec{v} / dt)_{s.w.} = - ( \eta \beta_{1} G M_{\odot} / ( Q'_{1} r^{2} ) )
( (\vec{v} \cdot \vec{e}_{1} / c ) \vec{e}_{1} + \vec{v} / c ); \eta
\approx 0.3 ]$, or some other nongravitational accelerations.

\acknowledgements{}
This paper was supported by VEGA grant No. 1/7067/20.

\end{document}